\documentclass[prl,twocolumn,nofootinbib,showpacs,superscriptaddress,preprintnumbers,amsmath,amssymb]{revtex4-1}

\usepackage[utf8]{inputenc}
\usepackage{graphicx,color,hyperref}
\usepackage{dcolumn}
\usepackage{bm}
\usepackage{amsmath}
\usepackage{amsthm}
\usepackage{tikz}
\usepackage{pgf,pgfplots}
\usepackage{mathrsfs}
\usepackage{amssymb}

\newcommand{\dg}{\dagger}
\newcommand{\der}{\partial}

\newcommand{\fr}{\frac}
\newcommand{\wed}{\wedge}

\newcommand\Lcal{\mathcal{L}}

\newcommand\e{\mathrm{e}}
\newcommand\m{\mathrm{m}}

\newcommand\p{\partial}
\newcommand\rd{{\rm d}}

\usepackage[normalem]{ulem}  

\renewcommand\sout{\bgroup \color{red} \ULdepth=-.5ex \ULset}

\newcommand{\im}{\mathrm{i}}

\begin{document}
\preprint{KEK-TH-2246, J-PARC-TH-221}
\title{
Counting Nambu-Goldstone modes of higher-form global symmetries
}

\author{Yoshimasa~Hidaka}
\email{hidaka@post.kek.jp}
\affiliation{KEK Theory Center, Tsukuba 305-0801, Japan}
\affiliation{Graduate University for Advanced Studies (Sokendai), Tsukuba 305-0801, Japan}
\affiliation{RIKEN iTHEMS, RIKEN, Wako 351-0198, Japan}

\author{Yuji~Hirono}
\email{yuji.hirono@gmail.com}
\affiliation{
Asia Pacific Center for Theoretical Physics, Pohang 37673, Korea
}
\affiliation{
Department of Physics, POSTECH, Pohang 37673, Korea
}

\author{Ryo~Yokokura}
\email{ryokokur@post.kek.jp}
\affiliation{KEK Theory Center, Tsukuba 305-0801, Japan}
\affiliation{Department of Physics \& Research and Education Center for Natural Sciences,
Keio University, Hiyoshi 4-1-1, Yokohama, Kanagawa 223-8521, Japan}

\date{\today}

\begin{abstract}
We discuss the counting of Nambu-Goldstone (NG) modes 
associated with the spontaneous breaking of
higher-form global symmetries.
Effective field theories of NG modes are developed 
based on symmetry breaking patterns, 
using a generalized coset construction for higher-form symmetries. 
We derive a formula of the number of gapless NG modes, 
which involves expectation values of the commutators of conserved charges, possibly of different degrees. 
\end{abstract}

\maketitle

 \setcounter{footnote}{0}%
\def\thefootnote{$*$\arabic{footnote}}%
   \def\@makefnmark{\hbox
       to\z@{$\m@th^{\@thefnmark}$\hss}}%

\emph{Introduction.}---%
Spontaneous symmetry breaking (SSB) 
is a common thread of modern physics 
running through various fields 
from condensed-matter to high-energy physics.
When continuous symmetries are
spontaneously broken,
gapless excitations appear
and they are called the Nambu-Goldstone (NG) modes \cite{Nambu:1961tp,Goldstone:1961eq,Goldstone:1962es} (see Ref.~\cite{Beekman:2019pmi} for a recent review).
Due to their gapless nature,
they dominate the low-energy physics. 
In relativistic systems,
there is a one-to-one correspondence
between a broken generator and a gapless mode.
However, the Lorentz symmetry is not present
in many physically-interesting situations,
especially in condensed-matter systems.
In the absence of Lorentz invariance,
the one-to-one correspondence no longer holds
\cite{Nielsen:1975hm, Miransky:2001tw,Schafer:2001bq,Nambu:2004}
and
the number of NG modes $N_{\rm NG}$ 
can be smaller
than the number of broken symmetry generators $ N_{\rm BS}$
\cite{Watanabe:2011ec, Watanabe:2012hr,Hidaka:2012ym,Takahashi:2014vua,Hayata:2014yga,Watanabe:2014fva},
\begin{equation}
  N_{\rm NG} = N_{\rm BS} - \frac{1}2 {\rm rank\,} \rho_{ab},
  \label{eq:ng-0}
\end{equation}
where $\rho_{ab}$ is the matrix of the expectation value of
charge commutators, $\rho_{ab} \propto \langle [\im Q_a, Q_b] \rangle$.

Recently, those symmetries
are understood as a part of a wider class of symmetries,
called higher-form symmetries~\cite{gaiotto2015generalized}
(see also \cite{Batista:2004sc,Nussinov:2008aiy,Nussinov:2006iva,Nussinov:2009zz,Nussinov:2011mz,Pantev:2005rh,Pantev:2005zs, Pantev:2005wj,Banks:2010zn,Hellerman:2010fv,Kapustin:2014gua,Sharpe:2015mja}).
The defining feature of higher-form symmetries is that charged objects under those symmetries
are extended:~The charged objects for a $p$-form symmetry are $p$-dimensional.
From this point of view, an ordinary symmetry
corresponds to a 0-form symmetry, whose charged objects
are point-like.
This concept provides us with a unifying perspective, 
and has been used in describing topological orders 
\cite{Wen:1989iv,Wen:1990zza,Wen:1991rp,gaiotto2015generalized,Wen:2018zux}, formulating the magnetohydrodynamics \cite{Grozdanov:2016tdf, Glorioso:2018kcp,Armas:2018ibg, Armas:2018atq}, 
and realizing a new class 
\cite{gaiotto2015generalized, Kapustin:2013uxa, Yoshida:2015cia} 
of Symmetry Protected Topological phases~\cite{PhysRevB.87.155114}. 
Similarly to the ordinary symmetries, 
when a continuous higher-form symmetry is spontaneously broken,
gapless excitations appear \cite{gaiotto2015generalized,Lake:2018dqm}.
In fact, photons are understood as the NG bosons associated with the spontaneous breaking of a $U(1)$ 1-form symmetry~\cite{Kovner:1992pu,gaiotto2015generalized}.

The question we would like to address
in this Letter is:~How many gapless NG modes appear when the higher-form symmetries are spontaneously broken?
To answer this, 
we develop effective field theories by extending the coset construction~\cite{Coleman:1969sm,Callan:1969sn},
which is widely used for ordinary symmetries. 
In the case of higher-form symmetries, 
the counting of the number of physical NG modes 
becomes involved because of the gauge degrees of freedom. 
We derive a generalized counting formula
of gapless NG modes for systems that are not necessarily 
Lorentz-invariant, 
which reduces to Eq.~\eqref{eq:ng-0} when all the symmetries are 0-form symmetries. 

\emph{SSB and coset construction.}---%
We consider the spontaneous breaking
of continuous internal symmetries
that can include higher-form symmetries
in $D$-dimensional Minkowski spacetime, $\mathbb R^{1,D-1}$.
We assume that the translational symmetry is unbroken.
We label each of the broken generators by $A$,
that generates a $p_A$-form symmetry.
For $p_A \ge 1$, the symmetry is always Abelian.
Let us denote the
the charged object
for the $p_A$-form symmetry
by $W_A(C_{p_A})$, which is supported
on a closed $p_A$-dimensional submanifold $C_{p_A}$.
In the case of a $U(1)$ symmetry,
the action of the symmetry is
\begin{equation}
  W_A(C_{p_A}) \mapsto
  e^{\im \alpha }   W_A(C_{p_A}) .
\end{equation}
The spontaneous breaking
of the symmetry $A$
is diagnosed
by
Coulomb or perimeter behavior
the vacuum expectation value,
\begin{equation}
  \langle
  W_A(C_{p_A})
\rangle
\sim 1
\quad\text{or}\quad
  \langle
    W_A(C_{p_A})
  \rangle
  \sim
  e^{ - T \, {\rm perimeter\,}[C_{p_A}]} ,
\label{200727.1517}
\end{equation}
where ${\rm perimeter\,}  [C_{p_A}]$
is the $p_A$-dimensional volume of $C_{p_A}$, and $T$ is the tension.

We would like to construct the low-energy effective field theory
based on the symmetry breaking patterns of higher-form symmetries.
The coset construction is a powerful technique 
to write down the action of the effective theory systematically
for ordinary symmetries~\cite{Coleman:1969sm,Callan:1969sn,Leutwyler:1993iq,Leutwyler:1993gf} and also for spacetime symmetries
\cite{Ivanov:1975zq,Low:2001bw,Nicolis:2013lma,Hidaka:2014fra,Delacretaz:2014jka}.
The basic building block for this method is the Maurer-Cartan 1-form.
To apply the coset construction to higher-form symmetries, we need the corresponding object. 
For a $p_A$-form symmetry breaking,
we define a generalized Maurer-Cartan form $f_A$,
which is a $(p_A+1)$-form,
by
\begin{equation}
  e^{\im\int_{X} f_A}
  :=
  W_A^\dag(C_{p_A}) W_A(C'_{p_A}),
  \label{eq:MCForm}
\end{equation}
where $W_A(C_{p_A}) = \exp(\im \int_{C_{p_A}} a_A)$ is
a generalization of coset variables,
and $X$ is a $(p_A+1)$-dimensional subspace such that $\p X = C_{p_A} \cup (- C'_{p_A})$.
For a 0-form symmetry, the left-hand side of Eq.~\eqref{eq:MCForm}
should be understood as a path-ordered product.
The generalized Maurer-Cartan form 
$f_A$ should satisfy the flatness condition $\rd f_A = 0$, 
since the right-hand side of Eq.~\eqref{eq:MCForm} is 
independent of the choice of 
the interpolating manifold $X$. 
The corresponding equation for a 0-form symmetry is the Maurer-Cartan equation. 
Since $f_A$ is a closed $(p_A+1)$-form,
it can be regarded as
the conserved current of the dual $(D-p_A-2)$-form symmetry, 
which is emergent in the symmetry-broken phase. 
The coset variable has a gauge redundancy
under $a_A\to a_A+ \rd \theta_{A}$, 
where $\theta_{A}$ is a $(p_A-1)$-form gauge parameter, 
because $a_A$ is integrated over a closed subspace $C_{p_A}$. 
A symmetry transformation shifts the field $a_A$
by a flat connection, $a_A \mapsto a_A + \lambda_A $\footnote{
For a 0-form symmetry,
$a_A$ is a 0-form
and the symmetry transformation is a constant shift, 
to the leading order in the number of fields. 
}.
Using the generalized Maurer-Cartan form 
as a building block, we can obtain effective actions 
by writing down the terms consistent with the
spacetime symmetry of the systems of interest. 
More details on this construction will be given elsewhere~\cite{counting:full}. 

\emph{Generalized counting rule.}---%
We here present the generalized 
counting rule of gapless NG modes associated with spontaneously broken higher-form symmetries. 
In Lorentz-invariant systems,
the number of gapless modes
for a broken generator $A$
of a $p_A$-form symmetry
is given by \cite{Hata:1980hn, Tokuoka:1982dd}
\begin{equation}
  \mathcal N_{D, A} =  {}_{D-2} C_{p_A} ,
\label{200723.1625}
\end{equation}
in $D$-dimensional spacetime.
In the absence of Lorentz invariance,
the number can be reduced. 
In this Letter, we show that the total number of gapless NG modes
is given by
\begin{equation}
  N_{\rm NG} =
  \sum_A \mathcal N_{D, A }
  -
  \frac{1}{2} {\rm rank} \, M_{\alpha\beta},
  \label{eq:gen-counting}
\end{equation}
where the summation
is over all the broken generators\footnote{
In the summation, 
one does not need to include 
the dual symmetry, since it does not exist in the UV 
and has rather emerged as a consequence of SSB. 
For example, 
in the Maxwell theory, 
the $U(1)_\m^{[1]}$ magnetic 1-form symmetry
is broken as well as $U(1)_\e^{[1]}$ symmetry, 
and there is a mixed 't~Hooft anomaly between them. 
The breaking pattern and the anomaly can be reproduced 
by introducing an NG mode of either $U(1)^{[1]}_{\rm e}$ or
$U(1)^{[1]}_{\rm m}$ symmetries. 
}.
Here,
$M_{\alpha\beta}$
is an anti-symmetric matrix,
whose definition will be given below.
Let us first define a quantity
proportional to the expectation
values of charge commutators,
\begin{equation}
  \begin{split}
  M_{AB} (V_A, V_B)
  &:=
    \frac{
    \langle
  \,
  [\im Q^{[p_A]}_A (V_A)
  ,
  Q^{[p_B]}_B (V_B) ]
  \,
  \rangle }
  {
{\rm vol \,
}
[ V_A \cap V_B ]
  }
  \end{split}.
\label{eq:qqm}
\end{equation}
Here, we denote
the conserved charge associated with the generator $A$ as
$Q^{[p_A]}_A(V_A)$,
which is supported
on a $(D-p_A-1)$-dimensional subspace $V_A$
located in the spatial slice $\Sigma$,
and ${\rm vol\,}[V]$ indicates
the volume of a subspace $V$.
To account for the independent choices of
subspaces $V_{A,B}$,
we shall consider the following matrix,
\begin{equation}
  M_{AB}^{i_1\cdots i_{p_A}, j_1 \cdots j_{p_B}}
  :=\lim_{
    \substack{
    V_{i_1 \cdots i_{p_A}}\to\infty \\
    V_{j_1 \cdots j_{p_B}}\to\infty
    }}
     M_{AB} (
       V_{i_1 \cdots i_{p_A}}  ,
       V_{j_1 \cdots j_{p_B}}
     ),
     \label{eq:m-lim}
\end{equation}
where indices of $V_{i_1 \cdots i_{p_A}}$
specify a $(D-p_A -1)$-dimensional plane
placed inside the $(D-1)$-dimensional
spatial manifold $\Sigma$.
The indices $i_1, \cdots$ only take
spatial ones
and is ordered, $i_1<i_2<\cdots < i_{p_A}$.
For example, in the case of a 1-form symmetry
in 4-spacetime dimensions, i.e., $D=4$ and $p_A=1$,
the symmetry generator is 2-dimensional plane
and $V_i$ represents the plane perpendicular to the $i$-axis for $i=x,y,z$.
In taking the limit,
we first take $V_{i_1 \cdots i_{p_A}}\to\infty$
and then take $V_{j_1 \cdots j_{p_B}}\to\infty$.
We collectively denote the indices
using the Greek letters 
as $\alpha := \bigl(A, (i_1, \cdots , i_{p_A})\bigr).$ 
With this notation,
the matrix~\eqref{eq:m-lim} is denoted as $M_{\alpha\beta}$,
that appears in the formula~\eqref{eq:gen-counting}.
$M_{\alpha\beta}$ is a real anti-symmetric\footnote{
  When the space-time symmetry is involved, there are examples where the charge commutator is not anti-symmetric \cite{Kobayashi:2014xua}.
} square matrix of dimension
$\sum_{A} {}_{D-1} C_{p_A}$.
Its rank is always even, 
and $({\rm rank} \, M_{\alpha\beta})/2$ is an integer. 
For a higher-form symmetry, a generator is supported on a closed subspace of dimension less than $D-1$. 
When we place them in the spatial slice $\Sigma$,
we have to specify how to place it,
which is the role of the indices $(i_1, \cdots, i_{p_A})$.
Thus, the matrix $M_{\alpha\beta}$ contains the information of the expectation values of charge commutators, including how to place them, and that determines the number of gapless modes.

The formula~\eqref{eq:gen-counting} is a natural generalization
of the one for 0-form symmetries \cite{Hidaka:2012ym,Watanabe:2012hr,Watanabe:2011ec}.
When all the symmetries are 0-form symmetries, $p_A=0$ for all $A$,
then $\sum_A \mathcal N_{D, A} =\sum_A 1
= N_{\rm BS}$,
and all the symmetry generators are
supported on a $(D-1)$-dimensional subspace,
$Q^{[0]}_{A}(V_{D-1})$.
Then, the formula~\eqref{eq:gen-counting}
reduces to the counting rule~\eqref{eq:ng-0}
for 0-form symmetries.

Let us make a comment on the dispersion relations.
For 0-form symmetries,
the NG modes are classified into type A and B, and they typically have linear and quadratic
dispersion relations, respectively.
When higher-form symmetries are involved,
it is still possible to classify the modes into type A and B, 
but it is not clear if we can 
associate this to the behavior of dispersion relations in general.
For example, whether the dispersion is linear or quadratic may depend on the direction of the propagation, 
as can be seen in an example discussed later [see Eq.~\eqref{eq:dispersion}].

\emph{Examples.}---%
To illustrate the coset construction and the counting rule,
let us confirm that the photons are NG modes
associated with a $U(1)$ 1-form symmetry,
which we denote by $U(1)^{[1]}_{\rm e}$.
When $U(1)^{[1]}_{\rm e}$ is spontaneously broken, 
the coset variable is defined on a loop $C$ 
and can parametrized as
$W(C) := e^{\im \int_C a }$,
which is nothing but the Wilson loop.
Since it is defined on a loop,
the field $a$ has a gauge redundancy,
$a \mapsto a  + \rd \theta_0$,
with a $U(1)$ 0-form parameter $e^{\im \theta_0}$. 
Therefore, we can identify $a$ as a $U(1)$ 1-form gauge field.
The $U(1)^{[1]}_{\rm e}$ transformation of $a$ is
locally given by $a \mapsto a +\lambda_1$ with a flat 1-form $\lambda_1$.
The generalized Maurer-Cartan form $f$ is defined through 
$
 e^{\im \int_{S} f} := W^\dg (C) W (C'),
$
where $C$ and $C'$ are loops,
and $S$ is a 2-dimensional surface
with $\der S = C' \cup(- C)$.
We can identify $f$ as $f= \rd a$, 
and the flatness condition $\rd f = 0$ is nothing but the Bianchi identity. 
In the Lorentz-invariant case, 
the gauge-invariant term with the lowest
number of derivatives is given by 
\begin{equation}
-\fr{1}{2e^2}
f \wed \star f,
\label{200723.1742}
\end{equation}
where $e$ is a coupling constant, and
$\star$ is the Hodge star operator.
This is nothing but the Maxwell theory.
There are ${}_{D-2}C_1$ gapless excitations 
in $D$ spacetime dimensions (two photons for $D=4$), 
which is consistent with the formula~\eqref{eq:gen-counting}.

As a second example, let us consider
photons in the presence of the gradient of 
the $\theta$ angle in $(3+1)$ dimensions~\cite{Yamamoto:2015maz},
whose 
Lagrangian
is given by
\begin{equation}
  -
  \frac{1}{2 e^2} 
f \wedge \star f
 +
\frac{c}{2} \theta f \wedge f,
\end{equation}
where $c$ is a constant.
The theory has $U(1)_\e^{[1]}$ symmetry as in the case of
the Maxwell theory, and
the corresponding conserved charge is identified as
$
  Q^{[1]}_\e(S) = \frac{1}{e^2}\int_{S} \star f
-
c\int_{S}\theta f.
$ 
This symmetry is spontaneously broken,
and $Q^{[1]}_\e(S)$ is a broken generator.
Due to the presence of the nonvanishing
$\rd \theta$,
the charge $Q^{[1]}_\e(S)$
becomes non-commutative,
\begin{equation}
 \langle
[
\im Q^{[1]}_\e (S_1) ,
Q^{[1]}_\e (S_2)
]\rangle
\propto \int_{S_1 \cap S_2} \rd\theta
\neq 0 ,
\end{equation}
where $S_1, S_2 \subset \Sigma$ are
2-dimensional surfaces
located inside the spatial manifold $\Sigma$.
Let us organize the charges as a vector,
$
  Q_i =
  \left(
    Q^{[1]}_\e (S_x),
    Q^{[1]}_\e (S_y),
    Q^{[1]}_\e (S_z)
  \right)^T ,
$
where $S_i$ is the surface perpendicular to $i$-axis. 
The charge commutator matrix is given by
\begin{equation}
  M_{ij}
  =
  \langle
   [\im Q_i, Q_j]
  \rangle
  \propto
  \epsilon_{ijk} \partial_k\theta,
\end{equation}
where the gradient $\partial_k\theta$ is constant.
The rank of this matrix is $2$, and
the number of gapless NG modes
via the formula~\eqref{eq:gen-counting}
is
\begin{equation}
  N_{\rm NG} =
  2
  - \frac{1}2 {\rm rank\, }M_{ij}  = 1.
\end{equation}
This coincides with the
result obtained by explicitly solving the equations of
motion (EOM).
There are one gapless and one gapped modes~\cite{Yamamoto:2015maz},
whose gap squared is given by
$\omega^2(k=0)  = C^2$.
Here, $C := |\bm C|$ is the length of the vector
$\bm C := -c e^2 \nabla \theta$.
The behavior of the dispersion relation depends on the direction of the momentum $\bm k$.
When the angle of $\bm k$ and $\bm C$ is  $\phi$,
the dispersion relation at small $k = |\bm k|$ is (see the Supplementary Material)
\begin{equation}
  \omega^2 (k) =
  \begin{cases}
(\sin^2\phi)\, k^2
 + \frac{\cos^4 \phi}{C^2} \, k^4
  + O(k^6),
  \\
  C^2 + (2 - \sin^2 \phi) k^2 + O(k^4).
  \end{cases}
  \label{eq:dispersion}
\end{equation}
The dispersion relation of the gapless mode
is quadratic, $\omega \sim k^2$,
only when $\sin \phi = 0$ ($\bm k$ is parallel to $\bm C $),
and is linear, $\omega \sim k$, for other angles.

A similar interpretation is possible 
for the system of neutral pions in the presence of background magnetic field~\cite{Sogabe:2019gif,Brauner:2017uiu}, where the commutators of 
0-form and 1-form symmetry generators 
acquire expectation values.

\emph{Derivation.}---%
Let us now give the derivation of the counting formula~\eqref{eq:gen-counting}.
For $p_A\geq1$, the higher-form symmetry is Abelian, 
so $f_A$ can be written as $\rd a_A$.
For a $0$-form non-Abelian symmetry, $f_A$ includes non-linear terms, that represent the interaction between NG modes. 
For the counting of the modes, it suffices to consider the linear order. Thus, we can also express the Maurer-Cartan form for $0$-form symmetry as $f_A=\rd a_A$, and we include indices of the Lie-algebra in $A$. 

For Lorentz invariant systems,
the kinetic terms are written,
to the lowest order in derivatives and fields, 
as 
\begin{equation}
  \Lcal_0 (\rd a_A)
  = -\frac{1}{2} F^2_{AB}f_A \wedge \star f_B
  = -\frac{1}{2} F^2_{AB}\rd a_A \wedge \star \rd a_B,
  \label{eq:L0}
\end{equation}
which is invariant under $ a_A \mapsto a_A + \lambda_A $.
Here, $F^2_{AB}=F_{CA}F_{CB}$ is a positive-definite symmetric matrix, and $F_{AB}$ corresponds to the decay-constant matrix.
The Lagrangian can also have topological terms of the form
$G_{AB} f_A\wedge f_B$,
where $G_{AB}$ are flat coefficients.
Those terms do not contribute to the EOM, so they can be dropped.
Since we can diagonalize $F^2_{AB}$,
it suffices to count the modes
when one $p_A$-form symmetry is spontaneously broken.
The EOM and the flatness condition\footnote{
The form $f_A$ is closed also for a non-Abelian 
0-form symmetry since we focus on the free part for the counting. 
} read 
\begin{equation}
  \rd^\dagger f_A = 0 ,
  \quad
  \rd f_A = 0 ,
  \label{eq:maxwell-bianchi}
\end{equation}
where $d^\dagger$ is the codifferential.
One can immediately see that
the field strength satisfies
$
  \Delta f_A = 0,
$
where $\Delta := \rd \rd^\dagger + \rd^\dagger \rd$ is
the Hodge Laplacian.
In the momentum space, the Hodge Laplacian
is given by $ - \omega^2 + k^2$.
Each component of the field strength satisfies
the wave equation, but
not all of them are independent.
The Bianchi identity and EOM
have
${}_{D-2} C_{p_A+1}$ and ${}_{D-2} C_{p_A-1}$
relations without time derivatives,
and these are the constraints.
The number of physical modes is given
by the number of components
of the field strength, ${}_{D} C_{p_A+1}$,
minus the number of constraints.
Since the physical modes consist of pairs of these degrees of freedom, we find
\begin{equation}
  {\mathcal N}_{D, p_A}
  =
  \frac{1}{2}( {}_{D} C_{p_A+1}
  -   {}_{D-2} C_{p_A+1}
  - {}_{D-2} C_{p_A-1})
   =
   {}_{D-2} C_{p_A} .
   \label{eq:n-d-p}
\end{equation}
Thus, the number of gapless modes
in the Lorentz invariant system
is given by the sum of this number over each generators,
\begin{equation}
  N_{\rm gapless} = \sum_A
  {\mathcal N}_{D, p_A} .
  \label{eq:n-gapless-rel}
\end{equation}

Now let us discuss the case without Lorentz invariance.
The breaking of the Lorentz invariance
means that there are special directions in spacetime.
We still assume that
the translational symmetry is not broken.
A key observation here is that
the absence of the Lorentz symmetry
can be represented as
the presence of external objects
placed in the spacetime.
This allows us to write additional terms 
with one derivative, 
\begin{equation}
 \Lcal_1 (a_A, f_A)
 :=
\frac{1}2 f_A \wedge  a_B \wedge \Omega_{AB} ,
\label{eq:l-omega}
\end{equation}
where $\Omega_{AB}$ is a flat $d$-form with $d := D - p_A - p_B - 1$.
The components of $\Omega_{AB}$ satisfy $\Omega_{AB} =- (-)^{p_Ap_B} \Omega_{BA}$
so that they are not a total derivative. 
Those terms~\eqref{eq:l-omega} 
are invariant under symmetry transformations 
only up to a total derivative,
and are reminiscent of the Wess-Zumino term~\cite{Wess:1971yu,Witten:1983tw}. 
We assume here that the external defect is
located {\it spatially},
by which we mean that $\Omega_{AB}$ does not involve $\rd x^0$.
Unless both $A$ and $B$ are both $0$-form symmetries, the external defect 
$\Omega_{AB}$ 
explicitly breaks the spatial rotation symmetry, which 
can lead to direction-dependent 
dispersion relations, as is seen in Eq.~\eqref{eq:dispersion}. 
The terms with two derivative 
can also be more generic compared to the relativistic case. 
Although such a deformation leads to more generic dispersion relations, 
it does not change the number of constraints,
so the number of physical gapless modes is unaffected. 
Thus, for the purpose of the counting of the modes, 
it suffices to use the following Lagrangian, 
\begin{equation}
  \Lcal = \Lcal_0 + \Lcal_1,
  \label{eq:lagrangian01}
\end{equation}
where $\Lcal_0$ is given by Eq.~\eqref{eq:L0}.

We can identify the existence of the terms~\eqref{eq:l-omega} 
with the non-vanishing expectation values of charge commutators. 
The Noether current associated with the generator
$A$ is given by 
\begin{equation}
  \star J_A
  = -F^2_{AB} \star f_B
  +  a_B \wedge
\Omega_{AB}.
  \label{eq:current-def}
\end{equation}
The symmetry generator is obtained
by integrating the current
over a submanifold $V_{D-p_A-1} \subset \Sigma$,
\begin{equation}
Q_A^{[p_A]} (V_{D-p_A-1})
 =
\int_{V_{D-p_A-1}} \star J_A.
\end{equation}
Since the action is quadratic, 
the current commutator
can be readily computed 
using the canonical commutation relations 
or the Ward-Takahashi identity as 
\begin{equation}
  \begin{split}
 \langle
 &
  [
  \im Q_A^{[p_A]} (V_{D-p_A-1}),
  \,\,
  Q_B^{[p_B]} (V_{D-p_B-1})
  ]
  \rangle  \\
  &
  \quad \quad \quad \quad \quad
  \propto 
   \int_{V_{D-p_A-1} \cap V_{D-p_B-1}}
   \Omega_{AB} .
\end{split}
\label{eq:q-q-omega}
\end{equation}

Now we are ready to discuss the number of 
physical gapless modes. 
The EOM 
of the Lagrangian~\eqref{eq:lagrangian01}
are written
in the form of current conservation,
$\rd \star J_A = 0$.
Let us write this with
gauge-invariant variables 
$ \tilde{f}_A = F_{AB} {f}_B$ as 
 \begin{equation}
  \rd \star \tilde{f}_A
  =   \tilde{f}_B \wedge  \widetilde{\Omega}_{AB} ,
  \label{eq:j-0-eom}
\end{equation}
where 
$\widetilde{\Omega}_{AB} := F_{AC}\Omega_{CD} F_{BD}$.
If we write out Eq.~\eqref{eq:j-0-eom}
explicitly with indices,
\begin{equation}
  \begin{split}
 & -\p^\alpha (\tilde{f}_A)_{ \mu_1 \cdots \mu_{p_A}\alpha} \\
&=
  \frac{\epsilon_{
    \mu_1 \cdots \mu_{p_A}}
    {}^
    {
    \nu_1 \cdots \nu_{p_B + 1}
    \rho_1 \cdots \rho_d
   }}{(p_B+1)!d!}
   (\widetilde\Omega_{AB})_{\rho_1 \cdots \rho_d}
   (\tilde{f}_B)_{\nu_1 \cdots \nu_{p_B+1}}.
  \end{split}
   \label{eq:eom-indices}
\end{equation}
In the relativistic case,
we have $\sum_A {}_{D-2} C_{p_A}$ gapless modes, 
but the presence of $\Omega_{AB}$ makes
a part of them gapped. 
We shall here count the number of {\it gapped} modes.
For this purpose, we write Eq.~\eqref{eq:eom-indices} in the energy/momentum space
and take $\bm k = \bm{0}$ and $\omega \neq 0$,
\begin{equation}
\begin{split}
- \im &\omega  \, (\tilde{f}_A)_{ i_1 \cdots i_{p_A}0} \\
  &=
  \epsilon_{
    i_1 \cdots i_{p_A}
  }{}^{
     j_1 \cdots j_{p_B }0
    k_1 \cdots k_d
   }
   (\widetilde\Omega_{AB})_{k_1 \cdots k_d}
   (\tilde{f}_B)_{ j_1 \cdots j_{p_B}0},
\end{split}
\end{equation}
where the Roman indices $i_1, \cdots ,j_1, \cdots, k_1, \cdots$ are spatial ones.
The sum of the indices is implicitly taken such that they satisfy $k_1<k_2<\cdots <k_d$, and $j_1<j_2<\cdots<j_{p_B}$, which removes the redundancy in the sum. 
Let us introduce a matrix notation
\begin{equation}
\widetilde{M}_{AB}^{i_1\cdots i_{p_A},j_1 \cdots j_{p_B}}
 =
 {\epsilon_{i_1 \cdots i_{p_A}}}^{
   j_1 \cdots j_{p_B}0k_1 \cdots k_d
   }
 (\widetilde\Omega_{AB})_{k_1 \cdots k_d}  .
\label{eq:omega-tilde-def}
\end{equation}
Then we have
\begin{equation}
 - \im \omega \, (\tilde{f}_A)_{ i_1 \cdots i_{p_A}0}
    =
    \widetilde M_{AB}^{i_1\cdots i_{p_A}, j_1 \cdots j_{p_B}}
    (\tilde{f}_B)_{ j_1 \cdots j_{p_B}0}.
\end{equation}
By using the short-hand notation,
$\alpha := (A, (i_1, \cdots , i_{p_A}))$, it can be further written as
$
- \im \omega \, \tilde{f}_\alpha =
\widetilde M_{\alpha \beta} \tilde{f}_\beta
$. 
The matrix $\widetilde M_{\alpha\beta}$ is antisymmetric,
and
it can be always transformed to the following form
by an orthogonal matrix $O$,
\begin{equation}
O \widetilde M O^T=
 \begin{pmatrix}
  \im \sigma_2 \lambda_1  &  & & & &  \\
   &  \ddots &  &  \\
    &   &\im \sigma_2 \lambda_m &  \\
&   &  &  & \bm 0
 \end{pmatrix},
\end{equation}
where $\sigma_2$ is the second Pauli matrix,
$m :=( {\rm rank\,} \widetilde M_{\alpha\beta} )/2$,
and $\lambda_i \neq 0$ for $i=1, \cdots, m$.
Each $2 \times 2$ sector in the upper-left part
gives a gapped mode.
For example,
\begin{equation}
  - \im
  \omega
  \begin{pmatrix}
    \tilde{f}_{\alpha=1} \\
    \tilde{f}_{\alpha=2}
  \end{pmatrix}
  =
  \begin{pmatrix}
    0 & \lambda_1 \\
    - \lambda_1  & 0
  \end{pmatrix}
  \begin{pmatrix}
    \tilde{f}_{\alpha=1} \\
    \tilde{f}_{\alpha=2}
  \end{pmatrix},
\end{equation}
represents a gapped excitation
whose gap is given by $\omega^2 = \lambda_1^2$.
Thus, the number of gapped modes is given by
\begin{equation}
  N_{\rm gapped} =
  \frac{1}{2} {\rm rank} \, \widetilde M_{\alpha\beta}
  =
  \frac{1}{2} {\rm rank} \, M_{\alpha\beta},
  \label{eq:n-gapped}
\end{equation}
where
$M_{\alpha\beta}$ is defined similarly to $\widetilde M_{\alpha\beta}$
by replacing $\widetilde \Omega_{AB}$ with $\Omega_{AB}$ in Eq.~\eqref{eq:omega-tilde-def},
and we used the fact that $F_{AB}$ is invertible.
The matrix $M_{\alpha\beta}$ is
nothing but the matrix \eqref{eq:m-lim}.
Since the presence of $\Lcal_1$ does not change the number of physical degrees of freedom, 
the number of {\it gapless} modes
in the absence of Lorentz invariance
is given by Eq.~\eqref{eq:n-gapless-rel}
minus Eq.~\eqref{eq:n-gapped},
which is
the formula~\eqref{eq:gen-counting}.

\begin{acknowledgments}
We thank Naoki Yamamoto and Noriyuki Sogabe for helpful discussions.
 Y. Hidaka is supported by JSPS KAKENHI Grant Numbers~17H06462 and 18H01211.   Y. Hirono is supported by the Korean Ministry of Education, Science and Technology, Gyeongsangbuk-do and Pohang City at the Asia Pacific Center for Theoretical Physics (APCTP) and by the National Research Foundation (NRF) funded by the Ministry of Science of Korea Grant Number 2020R1F1A1076267.
\end{acknowledgments}

\section{Supplementary material}

\subsection{
Photons under a constant $\theta$ gradient 
}

Here, we detail on the 
dispersion relation of a model of 
photons under a space-dependent $\theta$ angle. 
The Lagrangian of the system we consider is 
\begin{equation}
  \begin{split}
  \Lcal 
  &=   -
  \frac{1}{2 e^2}f \wedge \star f+\frac{c}{2}\theta f \wedge f 
  \\
  &=- \frac{1}{4e^2} f_{\mu\nu} f^{\mu\nu}d^4x
  - \frac{c}8 \theta\epsilon^{\mu\nu\alpha\beta} f_{\mu\nu}   f_{\alpha\beta}d^4x ,
\end{split}
\end{equation}
where $e$ is the coupling constant, $f_{\mu\nu} = \p_\mu a_\nu - \p_\nu a_\mu$ is the field strength with the photon field $a_\mu$, $c$ is a constant, and $\epsilon^{\mu\nu\rho\sigma}$ is the totally antisymmetric tensor with $\epsilon^{0123}=-1$.
The corresponding equations of motion are given, 
in the non-relativistic notation, by 
\begin{eqnarray}
  \nabla \cdot \bm e &=& - \bm C \cdot \bm b,  \\
 - \dot{\bm e}+  \nabla \times \bm b  
 &=&  \bm C \times \bm e, \\
  \nabla \cdot \bm b &=& 0, \\
   \dot{\bm b} + \nabla \times \bm e  &=& \bm 0 . 
\end{eqnarray}
where $\bm e=(f^{01},f^{02},f^{03})/e$ and $\bm b=(f_{23},f_{31},f_{12})/e$ are electric and magnetic fields, 
$\bm C = -c e^2\nabla \theta$ is taken to be a constant vector, and we used $\p_0 \theta = 0$. 
From those equations, 
we can derive the following equations with second-order derivatives, 
\begin{eqnarray}
  -  \ddot{\bm e} + \nabla^2 \bm e
- \nabla (\nabla \cdot \bm e)
- \bm C \times \dot{\bm e} &=& \bm 0 ,\\
-  \ddot {\bm b}
+ \nabla^2 \bm b 
+ \bm C (\nabla \cdot \bm e) 
- (\bm C \cdot \nabla) \bm e &=& \bm 0 . 
\end{eqnarray}
The corresponding equations in energy/momentum space are 
\begin{eqnarray}
(\omega^2-k^2) \bm e 
  + \bm k (\bm k \cdot \bm e)
  + \im \omega  \bm C \times {\bm e} &=& \bm 0 , \\
 (\omega^2 - k^2) \bm b 
 + \im \bm C (\bm k \cdot \bm e )
  -\im (\bm C \cdot \bm k) \bm e 
  &=& \bm 0, 
\end{eqnarray}
where $k:=|\bm k|$. 
Let us here use the basis 
$\{\hat{\bm k}, \bm \epsilon_{(1)}, \bm \epsilon_{(2)} \}$, 
where $\hat{\bm k}$ is the unit vector 
in the direction of $\bm k$, 
and two transverse unit vectors are defined by 
\begin{eqnarray}
  \bm \epsilon_{(1)} 
&:=& \frac{1}{\sin \phi}\hat{\bm k} \times \hat {\bm C} , 
\\
\bm \epsilon_{(2)} 
 &:=& \frac{1}{\sin \phi }
\left( \cos \phi \hat{\bm k} - \hat{\bm C} \right)
\propto \hat{\bm k} \times (\hat{\bm k} \times \hat{\bm C} ) ,  
\end{eqnarray}
where $\phi$ is the angle between $\bm C$ and $\bm k$, 
and $\hat{\bm C} := \bm C / C$ 
with $C := |\bm C|$. 
The set $\{\hat{\bm k}, \bm \epsilon_{(1)}, \bm \epsilon_{(2)} \}$ forms a right-handed triple. 
We later use relations, 
\begin{equation}
  \begin{split}
  &\bm C \cdot \hat{\bm k} = C \cos \phi, 
  \quad 
  \bm C \cdot \bm \epsilon_{(1)} = 0, 
  \quad 
  \bm C \cdot \bm \epsilon_{(2)} = - C \sin \phi,  \\
 & \bm C \times \hat{\bm k}
  = - C \sin \phi \bm \epsilon_{(1)}, 
  \quad 
  \bm C \times \bm \epsilon_{(1)} 
  = 
  C \left(
  \sin \phi \hat {\bm k } + \cos \phi \bm \epsilon_{(2)}
  \right), \\
&
  \bm C \times  \bm \epsilon_{(2)} 
  = - C\cos \phi \bm \epsilon_{(1)}. 
\end{split}
\end{equation}

Since the equations of motion for the electric fields are closed, let us use them to identify the dispersion relations. 
We decompose $\bm e$ 
with the basis $\{ \hat {\bm k}, \bm \epsilon_{(1)}, \bm \epsilon_{(2)}\}$ as 
\begin{equation}
  \bm e = 
  \chi_0 \, \hat{\bm k} 
  + 
  \chi_1 \bm \epsilon_{(1)}
  +
  \chi_2 \bm \epsilon_{(2)} . 
\end{equation}
Noting that 
\begin{equation}
  \begin{split}
  \bm C \times \bm e 
  &= 
  \bm C \times 
  \left(
    \chi_0 \, \hat{\bm k} 
    + 
    \chi_1 \bm \epsilon_{(1)}
    +
    \chi_2 \bm \epsilon_{(2)}
  \right)
\\    
&= 
- C \sin \phi \chi_0 \bm \epsilon_{(1)}
+ C \left(
  \sin \phi \hat {\bm k } + \cos \phi \bm \epsilon_{(2)}
  \right) \chi_1 
  \\
  &
  \quad      
 - C\cos \phi \bm \epsilon_{(1)} \chi_2   \\
 &= 
 C \sin \phi \chi_1 \hat {\bm k } 
- 
 C
 \left(
  \sin \phi \chi_0 + \cos \phi \chi_2 
 \right) \bm \epsilon_{(1)} \\
 & \quad 
 + C \cos \phi  \chi_1 \bm \epsilon_{(2)} , 
\end{split}
\end{equation}
we can write the 
equations of motion of the electric field as 
\begin{equation}
  \begin{split}
&
 (\omega^2 - k^2) \bm e 
    + k^2 \chi_0 \hat{\bm k} 
    \\
    &
    + \im \omega C
    \left(
     \sin \phi \chi_1 \hat {\bm k } 
     - 
     \left(
      \sin \phi \chi_0 + \cos \phi \chi_2 
     \right) \bm \epsilon_{(1)}
     +  \cos \phi  \chi_1 \bm \epsilon_{(2)} 
       \right) \\
&= 
(\omega^2 \chi_0 + \im \omega C \sin\phi \chi_1)
\hat {\bm k} \\
&
\quad + 
\left[
(\omega^2 - k^2) \chi_1 
- \im \omega C ( \sin \phi \chi_0 + \cos \phi \chi_2) 
\right]
\bm \epsilon_{(1)} 
\\
& \quad 
+ \left[
(\omega^2 - k^2) \chi_2 
+ \im \omega C \cos \phi  \chi_1)
\right]
\bm \epsilon_{(2)} \\
&=0 . 
  \end{split}
\end{equation}
Since the basis vectors are independent, 
we have 
\begin{equation}
  \begin{pmatrix}
    \omega^2 & \im \omega C \sin \phi & 0 \\
- \im \omega C \sin \phi  & \omega^2 - k^2 & 
- \im \omega C \cos \phi 
 \\
0 &
 \im \omega C \cos \phi 
& \omega^2 - k^2 
  \end{pmatrix}
  \begin{pmatrix}
    \chi_0 \\
    \chi_1 \\
    \chi_2 \\
  \end{pmatrix}
  = \bm 0 . 
\end{equation}
We can identify the dispersion relations 
of two physical modes as 
\begin{equation}
  \omega^2 (k) =
  \frac{1}2 
  \left(
    C^2 + 2 k^2 \pm \sqrt{ 
      C^4 + 2 C^2 k^2 (1+ \cos 2 \phi ) 
    }
  \right). 
\end{equation}
At small $k$, they behave as 
\begin{equation}
  \omega^2 (k) =
  \begin{cases}
  C^2 + (2 - \sin^2 \phi) k^2 + O(k^4)
    \\
 \sin^2 \phi \, k^2 
 + \frac{\cos^4 \phi}{C^2} \, k^4 
  + O(k^6)
  \end{cases} . 
\end{equation}
One mode is gapped and the other one is gapless. 
The dispersion relation of the gapless mode 
is quadratic, $\omega \sim k^2$, 
only when $\sin \phi = 0$, 
and is linear, $\omega \sim k$, for other angles.

\bibliographystyle{utphys}

\bibliography{refs}

\providecommand{\href}[2]{#2}\begingroup\raggedright\begin{thebibliography}{10}

\bibitem{Nambu:1961tp}
Y.~Nambu and G.~Jona-Lasinio, ``{Dynamical Model of Elementary Particles Based
  on an Analogy with Superconductivity. 1.},''
  \href{http://dx.doi.org/10.1103/PhysRev.122.345}{{\em Phys. Rev.} {\bfseries
  122} (1961) 345--358}.

\bibitem{Goldstone:1961eq}
J.~Goldstone, ``{Field Theories with Superconductor Solutions},''
  \href{http://dx.doi.org/10.1007/BF02812722}{{\em Nuovo Cim.} {\bfseries 19}
  (1961) 154--164}.

\bibitem{Goldstone:1962es}
J.~Goldstone, A.~Salam, and S.~Weinberg, ``{Broken Symmetries},''
  \href{http://dx.doi.org/10.1103/PhysRev.127.965}{{\em Phys. Rev.} {\bfseries
  127} (1962) 965--970}.

\bibitem{Beekman:2019pmi}
A.~J. Beekman, L.~Rademaker, and J.~van Wezel, ``{An Introduction to
  Spontaneous Symmetry Breaking},''
  \href{http://dx.doi.org/10.21468/SciPostPhysLectNotes.11}{{\em SciPost Phys.
  Lect. Notes} (2019) 11}, \href{http://arxiv.org/abs/1909.01820}{{\ttfamily
  arXiv:1909.01820 [hep-th]}}.
  \url{https://scipost.org/10.21468/SciPostPhysLectNotes.11}.

\bibitem{Nielsen:1975hm}
H.~B. Nielsen and S.~Chadha, ``{On How to Count Goldstone Bosons},''
\href{http://dx.doi.org/10.1016/0550-3213(76)90025-0}{{\em Nucl. Phys.}
  {\bfseries B105} (1976) 445}.

\bibitem{Miransky:2001tw}
V.~Miransky and I.~Shovkovy, ``{Spontaneous symmetry breaking with abnormal
  number of Nambu-Goldstone bosons and kaon condensate},''
  \href{http://dx.doi.org/10.1103/PhysRevLett.88.111601}{{\em Phys. Rev. Lett.}
  {\bfseries 88} (2002) 111601},
  \href{http://arxiv.org/abs/hep-ph/0108178}{{\ttfamily arXiv:hep-ph/0108178}}.

\bibitem{Schafer:2001bq}
T.~Schäfer, D.~Son, M.~A. Stephanov, D.~Toublan, and J.~Verbaarschot, ``{Kaon
  condensation and Goldstone's theorem},''
  \href{http://dx.doi.org/10.1016/S0370-2693(01)01265-5}{{\em Phys. Lett. B}
  {\bfseries 522} (2001) 67--75},
  \href{http://arxiv.org/abs/hep-ph/0108210}{{\ttfamily arXiv:hep-ph/0108210}}.

\bibitem{Nambu:2004}
Y.~Nambu, ``Spontaneous breaking of lie and current algebras,''
  \href{http://dx.doi.org/10.1023/B:JOSS.0000019827.74407.2d}{{\em J. Stat.
  Phys.} {\bfseries 115} no.~1, (2004) 7--17}.
  \url{https://doi.org/10.1023/B:JOSS.0000019827.74407.2d}.

\bibitem{Watanabe:2011ec}
H.~Watanabe and T.~Brauner, ``{On the number of Nambu-Goldstone bosons and its
  relation to charge densities},''
  \href{http://dx.doi.org/10.1103/PhysRevD.84.125013}{{\em Phys. Rev.}
  {\bfseries D 84} (2011) 125013},
\href{http://arxiv.org/abs/1109.6327}{{\ttfamily arXiv:1109.6327 [hep-ph]}}.

\bibitem{Watanabe:2012hr}
H.~Watanabe and H.~Murayama, ``{Unified Description of Nambu-Goldstone Bosons
  without Lorentz Invariance},''
  \href{http://dx.doi.org/10.1103/PhysRevLett.108.251602}{{\em Phys. Rev.
  Lett.} {\bfseries 108} (2012) 251602},
\href{http://arxiv.org/abs/1203.0609}{{\ttfamily arXiv:1203.0609 [hep-th]}}.

\bibitem{Hidaka:2012ym}
Y.~Hidaka, ``{Counting rule for Nambu-Goldstone modes in nonrelativistic
  systems},'' \href{http://dx.doi.org/10.1103/PhysRevLett.110.091601}{{\em
  Phys. Rev. Lett.} {\bfseries 110} (2013) 091601},
\href{http://arxiv.org/abs/1203.1494}{{\ttfamily arXiv:1203.1494 [hep-th]}}.

\bibitem{Takahashi:2014vua}
D.~A. Takahashi and M.~Nitta, ``{Counting rule of Nambu–Goldstone modes for
  internal and spacetime symmetries: Bogoliubov theory approach},''
  \href{http://dx.doi.org/10.1016/j.aop.2014.12.009}{{\em Annals Phys.}
  {\bfseries 354} (2015) 101--156},
\href{http://arxiv.org/abs/1404.7696}{{\ttfamily arXiv:1404.7696
  [cond-mat.quant-gas]}}.

\bibitem{Hayata:2014yga}
T.~Hayata and Y.~Hidaka, ``{Dispersion relations of Nambu-Goldstone modes at
  finite temperature and density},''
  \href{http://dx.doi.org/10.1103/PhysRevD.91.056006}{{\em Phys. Rev.}
  {\bfseries D91} (2015) 056006},
\href{http://arxiv.org/abs/1406.6271}{{\ttfamily arXiv:1406.6271 [hep-th]}}.

\bibitem{Watanabe:2014fva}
H.~Watanabe and H.~Murayama, ``{Effective Lagrangian for Nonrelativistic
  Systems},'' \href{http://dx.doi.org/10.1103/PhysRevX.4.031057}{{\em Phys.
  Rev.} {\bfseries X 4} no.~3, (2014) 031057},
\href{http://arxiv.org/abs/1402.7066}{{\ttfamily arXiv:1402.7066 [hep-th]}}.

\bibitem{gaiotto2015generalized}
D.~Gaiotto, A.~Kapustin, N.~Seiberg, and B.~Willett, ``Generalized global
  symmetries,'' {\em J. High Energy Phys.} {\bfseries 2015} no.~2, (2015) 172.

\bibitem{Batista:2004sc}
C.~D. Batista and Z.~Nussinov, ``{Generalized Elitzur's theorem and dimensional
  reduction},'' \href{http://dx.doi.org/10.1103/PhysRevB.72.045137}{{\em Phys.
  Rev. B} {\bfseries 72} (2005) 045137},
  \href{http://arxiv.org/abs/cond-mat/0410599}{{\ttfamily
  arXiv:cond-mat/0410599}}.

\bibitem{Nussinov:2008aiy}
Z.~Nussinov and G.~Ortiz, ``{Topological Quantum Order: a new Paradigm in the
  Physics of Matter},''
  \href{http://dx.doi.org/10.1142/9789812779885\_0053}{{\em Ser. Adv. Quant.
  Many Body Theor.} {\bfseries 11} (2008) 423--430}.

\bibitem{Nussinov:2006iva}
Z.~Nussinov and G.~Ortiz, ``{Sufficient symmetry conditions for Topological
  Quantum Order},'' \href{http://dx.doi.org/10.1073/pnas.0803726105}{{\em Proc.
  Nat. Acad. Sci.} {\bfseries 106} (2009) 16944--16949},
  \href{http://arxiv.org/abs/cond-mat/0605316}{{\ttfamily
  arXiv:cond-mat/0605316}}.

\bibitem{Nussinov:2009zz}
Z.~Nussinov and G.~Ortiz, ``{A symmetry principle for topological quantum
  order},'' \href{http://dx.doi.org/10.1016/j.aop.2008.11.002}{{\em Annals
  Phys.} {\bfseries 324} (2009) 977--1057},
  \href{http://arxiv.org/abs/cond-mat/0702377}{{\ttfamily
  arXiv:cond-mat/0702377}}.

\bibitem{Nussinov:2011mz}
Z.~Nussinov, G.~Ortiz, and E.~Cobanera, ``{Effective and exact holographies
  from symmetries and dualities},''
  \href{http://dx.doi.org/10.1016/j.aop.2012.07.001}{{\em Annals Phys.}
  {\bfseries 327} (2012) 2491--2521},
  \href{http://arxiv.org/abs/1110.2179}{{\ttfamily arXiv:1110.2179
  [cond-mat.stat-mech]}}.

\bibitem{Pantev:2005rh}
T.~Pantev and E.~Sharpe, ``{Notes on gauging noneffective group actions},''
  \href{http://arxiv.org/abs/hep-th/0502027}{{\ttfamily arXiv:hep-th/0502027}}.

\bibitem{Pantev:2005zs}
T.~Pantev and E.~Sharpe, ``{GLSM's for Gerbes (and other toric stacks)},''
  \href{http://dx.doi.org/10.4310/ATMP.2006.v10.n1.a4}{{\em Adv. Theor. Math.
  Phys.} {\bfseries 10} no.~1, (2006) 77--121},
  \href{http://arxiv.org/abs/hep-th/0502053}{{\ttfamily arXiv:hep-th/0502053}}.

\bibitem{Pantev:2005wj}
T.~Pantev and E.~Sharpe, ``{String compactifications on Calabi-Yau stacks},''
  \href{http://dx.doi.org/10.1016/j.nuclphysb.2005.10.035}{{\em Nucl. Phys. B}
  {\bfseries 733} (2006) 233--296},
  \href{http://arxiv.org/abs/hep-th/0502044}{{\ttfamily arXiv:hep-th/0502044}}.

\bibitem{Banks:2010zn}
T.~Banks and N.~Seiberg, ``{Symmetries and Strings in Field Theory and
  Gravity},'' \href{http://dx.doi.org/10.1103/PhysRevD.83.084019}{{\em Phys.
  Rev. D} {\bfseries 83} (2011) 084019},
  \href{http://arxiv.org/abs/1011.5120}{{\ttfamily arXiv:1011.5120 [hep-th]}}.

\bibitem{Hellerman:2010fv}
S.~Hellerman and E.~Sharpe, ``{Sums over topological sectors and quantization
  of Fayet-Iliopoulos parameters},''
  \href{http://dx.doi.org/10.4310/ATMP.2011.v15.n4.a7}{{\em Adv. Theor. Math.
  Phys.} {\bfseries 15} (2011) 1141--1199},
  \href{http://arxiv.org/abs/1012.5999}{{\ttfamily arXiv:1012.5999 [hep-th]}}.

\bibitem{Kapustin:2014gua}
A.~Kapustin and N.~Seiberg, ``{Coupling a QFT to a TQFT and Duality},''
  \href{http://dx.doi.org/10.1007/JHEP04(2014)001}{{\em J. High Energy Phys.}
  {\bfseries 04} (2014) 001},
\href{http://arxiv.org/abs/1401.0740}{{\ttfamily arXiv:1401.0740 [hep-th]}}.

\bibitem{Sharpe:2015mja}
E.~Sharpe, ``{Notes on generalized global symmetries in QFT},''
  \href{http://dx.doi.org/10.1002/prop.201500048}{{\em Fortsch. Phys.}
  {\bfseries 63} (2015) 659--682},
  \href{http://arxiv.org/abs/1508.04770}{{\ttfamily arXiv:1508.04770
  [hep-th]}}.

\bibitem{Wen:1989iv}
X.~Wen, ``{Topological Order in Rigid States},''
  \href{http://dx.doi.org/10.1142/S0217979290000139}{{\em Int. J. Mod. Phys. B}
  {\bfseries 4} (1990) 239}.

\bibitem{Wen:1990zza}
X.~Wen and Q.~Niu, ``{Ground-state degeneracy of the fractional quantum Hall
  states in the presence of a random potential and on high-genus Riemann
  surfaces},'' \href{http://dx.doi.org/10.1103/PhysRevB.41.9377}{{\em Phys.
  Rev. B} {\bfseries 41} (1990) 9377--9396}.

\bibitem{Wen:1991rp}
X.-G. Wen, ``{Topological orders and Chern-Simons theory in strongly correlated
  quantum liquid},'' \href{http://dx.doi.org/10.1142/S0217979291001541}{{\em
  Int. J. Mod. Phys. B} {\bfseries 5} (1991) 1641--1648}.

\bibitem{Wen:2018zux}
X.-G. Wen, ``{Emergent anomalous higher symmetries from topological order and
  from dynamical electromagnetic field in condensed matter systems},''
  \href{http://dx.doi.org/10.1103/PhysRevB.99.205139}{{\em Phys. Rev. B}
  {\bfseries 99} no.~20, (2019) 205139},
  \href{http://arxiv.org/abs/1812.02517}{{\ttfamily arXiv:1812.02517
  [cond-mat.str-el]}}.

\bibitem{Grozdanov:2016tdf}
S.~Grozdanov, D.~M. Hofman, and N.~Iqbal, ``{Generalized global symmetries and
  dissipative magnetohydrodynamics},''
  \href{http://dx.doi.org/10.1103/PhysRevD.95.096003}{{\em Phys. Rev. D}
  {\bfseries 95} no.~9, (2017) 096003},
  \href{http://arxiv.org/abs/1610.07392}{{\ttfamily arXiv:1610.07392
  [hep-th]}}.

\bibitem{Glorioso:2018kcp}
P.~Glorioso and D.~T. Son, ``{Effective field theory of magnetohydrodynamics
  from generalized global symmetries},''
  \href{http://arxiv.org/abs/1811.04879}{{\ttfamily arXiv:1811.04879
  [hep-th]}}.

\bibitem{Armas:2018ibg}
J.~Armas, J.~Gath, A.~Jain, and A.~V. Pedersen, ``{Dissipative hydrodynamics
  with higher-form symmetry},''
  \href{http://dx.doi.org/10.1007/JHEP05(2018)192}{{\em J. High Energy Phys.}
  {\bfseries 05} (2018) 192}, \href{http://arxiv.org/abs/1803.00991}{{\ttfamily
  arXiv:1803.00991 [hep-th]}}.

\bibitem{Armas:2018atq}
J.~Armas and A.~Jain, ``{Magnetohydrodynamics as superfluidity},''
  \href{http://dx.doi.org/10.1103/PhysRevLett.122.141603}{{\em Phys. Rev.
  Lett.} {\bfseries 122} no.~14, (2019) 141603},
  \href{http://arxiv.org/abs/1808.01939}{{\ttfamily arXiv:1808.01939
  [hep-th]}}.

\bibitem{Kapustin:2013uxa}
A.~Kapustin and R.~Thorngren, ``{Higher symmetry and gapped phases of gauge
  theories},'' \href{http://arxiv.org/abs/1309.4721}{{\ttfamily arXiv:1309.4721
  [hep-th]}}.

\bibitem{Yoshida:2015cia}
B.~Yoshida, ``{Topological phases with generalized global symmetries},''
  \href{http://dx.doi.org/10.1103/PhysRevB.93.155131}{{\em Phys. Rev. B}
  {\bfseries 93} no.~15, (2016) 155131},
  \href{http://arxiv.org/abs/1508.03468}{{\ttfamily arXiv:1508.03468
  [cond-mat.str-el]}}.

\bibitem{PhysRevB.87.155114}
X.~Chen, Z.-C. Gu, Z.-X. Liu, and X.-G. Wen, ``Symmetry protected topological
  orders and the group cohomology of their symmetry group,''
  \href{http://dx.doi.org/10.1103/PhysRevB.87.155114}{{\em Phys. Rev. B}
  {\bfseries 87} (Apr, 2013) 155114}.
  \url{https://link.aps.org/doi/10.1103/PhysRevB.87.155114}.

\bibitem{Lake:2018dqm}
E.~Lake, ``{Higher-form symmetries and spontaneous symmetry breaking},''
  \href{http://arxiv.org/abs/1802.07747}{{\ttfamily arXiv:1802.07747
  [hep-th]}}.

\bibitem{Kovner:1992pu}
A.~Kovner and B.~Rosenstein, ``{New look at QED in four-dimensions: The Photon
  as a Goldstone boson and the topological interpretation of electric
  charge},'' \href{http://dx.doi.org/10.1103/PhysRevD.49.5571}{{\em Phys. Rev.
  D} {\bfseries 49} (1994) 5571--5581},
\href{http://arxiv.org/abs/hep-th/9210154}{{\ttfamily arXiv:hep-th/9210154
  [hep-th]}}.

\bibitem{Coleman:1969sm}
S.~R. Coleman, J.~Wess, and B.~Zumino, ``{Structure of phenomenological
  Lagrangians. 1.},''
\href{http://dx.doi.org/10.1103/PhysRev.177.2239}{{\em Phys. Rev.} {\bfseries
  177} (1969) 2239--2247}.

\bibitem{Callan:1969sn}
J.~Callan, Curtis~G., S.~R. Coleman, J.~Wess, and B.~Zumino, ``{Structure of
  phenomenological Lagrangians. 2.},''
\href{http://dx.doi.org/10.1103/PhysRev.177.2247}{{\em Phys. Rev.} {\bfseries
  177} (1969) 2247--2250}.

\bibitem{Leutwyler:1993iq}
H.~Leutwyler, ``{On the foundations of chiral perturbation theory},''
  \href{http://dx.doi.org/10.1006/aphy.1994.1094}{{\em Annals Phys.} {\bfseries
  235} (1994) 165--203}, \href{http://arxiv.org/abs/hep-ph/9311274}{{\ttfamily
  arXiv:hep-ph/9311274}}.

\bibitem{Leutwyler:1993gf}
H.~Leutwyler, ``{Nonrelativistic effective Lagrangians},''
  \href{http://dx.doi.org/10.1103/PhysRevD.49.3033}{{\em Phys. Rev. D}
  {\bfseries 49} (1994) 3033--3043},
  \href{http://arxiv.org/abs/hep-ph/9311264}{{\ttfamily arXiv:hep-ph/9311264}}.

\bibitem{Ivanov:1975zq}
E.~Ivanov and V.~Ogievetsky, ``{The Inverse Higgs Phenomenon in Nonlinear
  Realizations},''
{\em Teor. Mat. Fiz.} {\bfseries 25} (1975) 164--177.

\bibitem{Low:2001bw}
I.~Low and A.~V. Manohar, ``{Spontaneously broken space-time symmetries and
  Goldstone's theorem},''
  \href{http://dx.doi.org/10.1103/PhysRevLett.88.101602}{{\em Phys. Rev. Lett.}
  {\bfseries 88} (2002) 101602},
\href{http://arxiv.org/abs/hep-th/0110285}{{\ttfamily arXiv:hep-th/0110285
  [hep-th]}}.

\bibitem{Nicolis:2013lma}
A.~Nicolis, R.~Penco, and R.~A. Rosen, ``{Relativistic Fluids, Superfluids,
  Solids and Supersolids from a Coset Construction},''
  \href{http://dx.doi.org/10.1103/PhysRevD.89.045002}{{\em Phys. Rev. D}
  {\bfseries 89} no.~4, (2014) 045002},
  \href{http://arxiv.org/abs/1307.0517}{{\ttfamily arXiv:1307.0517 [hep-th]}}.

\bibitem{Hidaka:2014fra}
Y.~Hidaka, T.~Noumi, and G.~Shiu, ``{Effective field theory for spacetime
  symmetry breaking},''
  \href{http://dx.doi.org/10.1103/PhysRevD.92.045020}{{\em Phys. Rev. D}
  {\bfseries 92} no.~4, (2015) 045020},
  \href{http://arxiv.org/abs/1412.5601}{{\ttfamily arXiv:1412.5601 [hep-th]}}.

\bibitem{Delacretaz:2014jka}
L.~V. Delacrétaz, A.~Nicolis, R.~Penco, and R.~A. Rosen, ``{Wess-Zumino Terms
  for Relativistic Fluids, Superfluids, Solids, and Supersolids},''
  \href{http://dx.doi.org/10.1103/PhysRevLett.114.091601}{{\em Phys. Rev.
  Lett.} {\bfseries 114} no.~9, (2015) 091601},
  \href{http://arxiv.org/abs/1403.6509}{{\ttfamily arXiv:1403.6509 [hep-th]}}.

\bibitem{counting:full}
Y.~Hidaka, Y.~Hirono, and R.~Yokokura {\em in preparation} .

\bibitem{Hata:1980hn}
H.~Hata, T.~Kugo, and N.~Ohta, ``{Skew Symmetric Tensor Gauge Field Theory
  Dynamically Realized in {QCD} U(1) Channel},''
  \href{http://dx.doi.org/10.1016/0550-3213(81)90170-X}{{\em Nucl. Phys. B}
  {\bfseries 178} (1981) 527--544}.

\bibitem{Tokuoka:1982dd}
Z.~Tokuoka, ``{A Classification of Masless Totally Antisymmetric tensor field
  in arbitrary dimensional space-time},''
  \href{http://dx.doi.org/10.1016/0375-9601(82)90005-6}{{\em Phys. Lett. A}
  {\bfseries 87} (1982) 215--219}.

\bibitem{Kobayashi:2014xua}
M.~Kobayashi and M.~Nitta, ``{Nonrelativistic Nambu-Goldstone Modes Associated
  with Spontaneously Broken Space-Time and Internal Symmetries},''
  \href{http://dx.doi.org/10.1103/PhysRevLett.113.120403}{{\em Phys. Rev.
  Lett.} {\bfseries 113} no.~12, (2014) 120403},
  \href{http://arxiv.org/abs/1402.6826}{{\ttfamily arXiv:1402.6826 [hep-th]}}.

\bibitem{Yamamoto:2015maz}
N.~Yamamoto, ``{Axion electrodynamics and nonrelativistic photons in nuclear
  and quark matter},'' \href{http://dx.doi.org/10.1103/PhysRevD.93.085036}{{\em
  Phys. Rev. D} {\bfseries 93} no.~8, (2016) 085036},
  \href{http://arxiv.org/abs/1512.05668}{{\ttfamily arXiv:1512.05668
  [hep-th]}}.

\bibitem{Sogabe:2019gif}
N.~Sogabe and N.~Yamamoto, ``{Triangle Anomalies and Nonrelativistic
  Nambu-Goldstone Modes of Generalized Global Symmetries},''
  \href{http://dx.doi.org/10.1103/PhysRevD.99.125003}{{\em Phys. Rev. D}
  {\bfseries 99} no.~12, (2019) 125003},
\href{http://arxiv.org/abs/1903.02846}{{\ttfamily arXiv:1903.02846 [hep-th]}}.

\bibitem{Brauner:2017uiu}
T.~Brauner and S.~V. Kadam, ``{Anomalous low-temperature thermodynamics of QCD
  in strong magnetic fields},''
  \href{http://dx.doi.org/10.1007/JHEP11(2017)103}{{\em J. High Energy Phys.}
  {\bfseries 11} (2017) 103}, \href{http://arxiv.org/abs/1706.04514}{{\ttfamily
  arXiv:1706.04514 [hep-ph]}}.

\bibitem{Wess:1971yu}
J.~Wess and B.~Zumino, ``{Consequences of anomalous Ward identities},''
  \href{http://dx.doi.org/10.1016/0370-2693(71)90582-X}{{\em Phys. Lett. B}
  {\bfseries 37} (1971) 95--97}.

\bibitem{Witten:1983tw}
E.~Witten, ``{Global Aspects of Current Algebra},''
  \href{http://dx.doi.org/10.1016/0550-3213(83)90063-9}{{\em Nucl. Phys. B}
  {\bfseries 223} (1983) 422--432}.

\end{thebibliography}\endgroup

\end{document}